# Electro-optic frequency combs for rapid interrogation in cavity optomechanics

D. A. Long,[1,*] B. J. Reschovsky,[1] F. Zhou,[1] Y. Bao,[1] T. W. LeBrun,[1] and J. J. Gorman[1,*]

[1]National Institute of Standards and Technology, 100 Bureau Dr, Gaithersburg, MD 20899
*Corresponding authors: David.long@nist.gov, Gorman@nist.gov

**Electro-optic frequency combs were employed to rapidly interrogate an optomechanical sensor, demonstrating spectral resolution substantially exceeding that possible with a mode-locked frequency comb. Frequency combs were generated using an integrated-circuit-based direct digital synthesizer and utilized in a self-heterodyne configuration. Unlike approaches based upon laser locking, the present approach allows rapid, parallel measurements of full optical cavity modes, large dynamic range of sensor displacement, and acquisition across a wide frequency range between DC and 500 kHz. In addition to being well suited to measurements of cavity optomechanical sensors, this optical frequency comb-based approach can be utilized for interrogation in a wide range of physical and chemical sensors.**

The measurement of optical resonance frequencies of cavity modes is essential in almost all experiments with cavity optomechanical systems [1] and is generally achieved using an optical readout method based on laser frequency locking [2-7]. These measurements are used to determine the displacement of mechanical resonators [8], changes in a cavity's effective refractive index [9], and to investigate dispersive or dissipative optomechanical interactions [10]. While laser frequency locking is widely used for laser stabilization in macroscopic systems [2-4, 11, 12], it is less effective for the readout of micro- or nanoscale cavity optomechanical systems. Changes in cavity length due to the motion of an optomechanical resonator can cause frequency shifts that are large compared to the cavity linewidth, requiring wide frequency tuning of the locked laser. In addition, this frequency tuning must have high bandwidth in many cases in order to maintain the lock, such as when the optomechanical resonator has both high vibration amplitude and a high resonance frequency.

The combination of wide frequency tuning and high tuning bandwidth is not found in most stable single-frequency lasers. For example, external cavity diode lasers (ECDLs) may have sufficient tuning range but the piezoelectric actuators used to tune the wavelength typically have bandwidths well below the mechanical resonance frequencies found in many cavity optomechanical systems [8, 10, 13, 14]. Also, although current tuning can provide high bandwidth in these lasers, it offers insufficient tuning range.

In certain circumstances, fast Pound-Drever-Hall (PDH) [2] laser locking with large tuning ranges can be achieved using external modulators, but these techniques have other challenges, such as the presence of extraneous sidebands or the need for precise stabilization of multiple bias voltages [5-7]. In addition, the high-gain, large-bandwidth controllers that are required amplify electronic noise over a large frequency band and contribute to readout noise. Finally, the broad linewidths of microcavity optical resonances (generally hundreds of MHz or more) requires large modulation frequencies, adding to the challenges of PDH locking.

Given these limitations in laser technology, optical cavity readout with laser frequency locking generally results in low feedback bandwidth or low laser tuning range, or both. This is particularly problematic for optomechanical sensors, where large range and bandwidth are essential for operation, so new readout methods that can meet these performance requirements are essential.

In this Letter, we present a new method for cavity readout that does not require laser locking, feedback control, or precision frequency tuning of the laser. An optical frequency comb generated with an electro-optic phase modulator is used to detect the full spectrum of a single resonance of an optical cavity within an optomechanical system. By sampling this spectrum at a high rate, the center frequency of the cavity resonance can be measured as a function of time, thereby providing the change in length of the cavity. Also, this method avoids the complexity and added controller noise of a fast-feedback system. Finally, very high dynamic range can be achieved by generating a wide frequency comb and the measurement range is limited only by the data acquisition and photodetector bandwidth, which can easily reach many GHz or more.

Results that were measured on an integrated cavity optomechanical sensor with the electro-optic frequency comb readout method are presented to demonstrate the effectiveness of the approach, including dynamic range, linearity, mechanical ring-down, noise floor, and cavity stability tests. Further, we note that the optical approach described herein is readily applicable to a wide range of optomechanical systems including many other types of physical and chemical sensors.

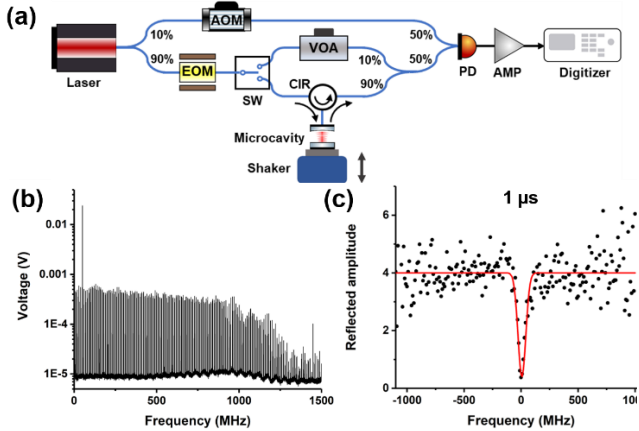

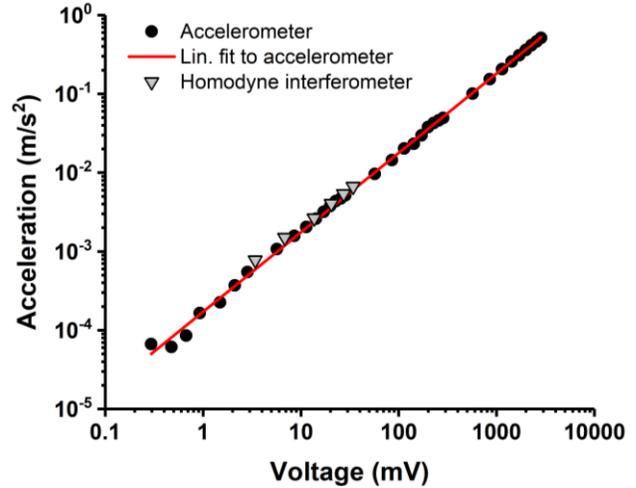

**Fig. 1.** (a) Diagram of the electro-optic frequency comb readout method. The laser source was a low-noise distributed feedback fiber laser. EOM: electro-optic phase modulator, AOM: acousto-optic modulator, SW: mechanical optical switch, CIR.: optical circulator, VOA: variable attenuator, PD: photodetector, AMP: amplifier. (b) Optical frequency comb recorded as the average of one hundred 30 kS records. The carrier tone can be seen at 51 MHz as well as the 10 MHz spaced optical frequency comb. (c) Measurement of a normalized optical cavity mode recorded in 1 μs and the corresponding Gaussian fit.

**Fig. 2.** Root-mean-square acceleration as a function of piezoelectric shaker table drive voltage for a 4 kHz drive frequency. In addition to the comb-based accelerometer measurements, independent measurements based upon a homodyne Michelson interferometer are shown. Over the voltage range of the homodyne interferometer measurements, the slopes of linear fits to the two measurements were within 6 %.

The implementation of the electro-optic frequency comb readout method is described in Fig. 1a. We generate a frequency comb by driving an electro-optic phase modulator with a repeating linear frequency chirp produced by a direct digital synthesizer (DDS) integrated circuit. The use of a constant amplitude, linear frequency chirp has been shown to be a nearly ideal approach for the generation of ultraflat frequency combs [15, 16]. In addition, this approach allows for combs whose properties can be controlled in an agile, digital fashion with comb tooth spacings that can be set over six orders of magnitude between hundreds of Hz and hundreds of MHz [17].

As this comb generation approach has been previously described for atomic spectroscopy applications [17], we will only briefly outline it here. The DDS generates a train of constant amplitude, linearly chirped waveforms that can span from 10 MHz to 1200 MHz with a widely tunable repetition period that was selected to be 100 ns in the presented measurements. This configuration produces an optical frequency comb that is centered at the carrier frequency of the laser and spans 2.4 GHz in width with a spacing of 10 MHz (i.e., the inverse of the repetition period). We note that the comb span could readily be increased through the use of either cascaded modulators [18] or nonlinear broadening [19]. The optical frequency comb is sent to the optomechanical sensor through a fiber-optic circulator and observed in reflection.

A self-heterodyne architecture [16, 20] is employed to down-convert the reflected optical frequency comb from the sensor into the radio frequency domain. A second optical path serves as the local oscillator, which is combined on a photodiode with the optical frequency comb reflected from the optomechanical cavity. An acousto-optic modulator in the local oscillator path shifts the carrier tone by 51 MHz to ensure that positive- and negative-order comb teeth occur at unique frequencies in the radiofrequency domain. An optical switch is employed to normalize the resulting measurements by a spectrum recorded when bypassing the sensor. This normalization, which can be performed as infrequently as once per day, addresses minor deviations from a flat optical comb and electrical frequency-response curve. In order to avoid any potential damage to the cavity optomechanical sensor, the total optical power incident on the device was limited to a few hundred μW.

The electro-optic frequency comb readout method is ideally suited to dynamic, high amplitude changes in the cavity length where laser locking approaches are generally precluded. In order to demonstrate the capabilities of the method, it was applied to an integrated cavity optomechanical accelerometer. The optomechanical component of the accelerometer is composed of a mechanical resonator and silicon concave micromirror, both with high-reflectivity mirror coatings. The resonator and micromirror form a hemispherical optical cavity with a $TEM_{00}$ mode that has a finesse of 5430, a cavity length of 375 μm, and a full-width at half-maximum linewidth of 73.7 MHz. The mechanical resonator has well-separated vibrational modes with a fundamental resonance located at 9.8 kHz. The accelerometer was packaged in a stainless-steel mount that facilitates fiber coupling of light into and out of the cavity and mounting to commercial shaker tables. More details on the design and fabrication of this accelerometer can be found in Ref. [21].

A typical electro-optic frequency comb after down conversion is shown in Fig. 1b. Continuous temporal interferograms with a length of 0.5 s were acquired at $3\times10^9$ samples per second and divided into 1 μs sub-interferograms to be fast Fourier transformed (FFT) and normalized to generate reflection spectra of an individual sensor cavity mode. The resulting 500,000 cavity mode spectra were then individually fit using a Gaussian profile as shown in Fig. 1 (c). See the Supplement 1 for more details on the data processing procedure.

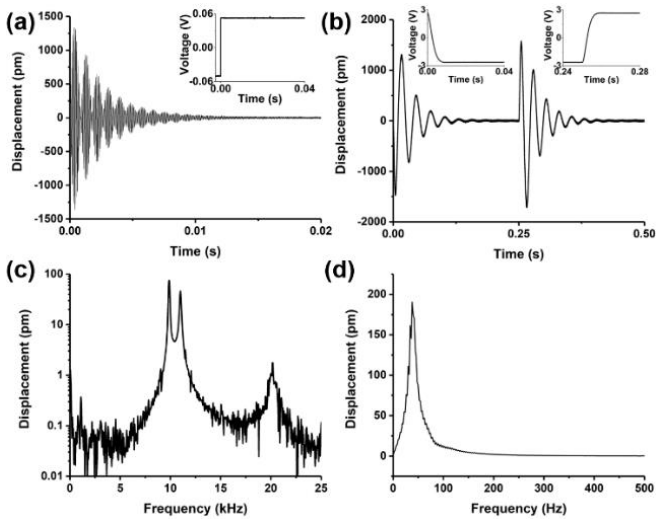

**Fig. 3.** (a) Displacement of the optical cavity within the accelerometer due to an unfiltered square wave excitation to the piezoelectric shaker (see inset). (b) Displacement of the mechanical resonator in the accelerometer due to a low frequency excitation via a square wave that has been low-pass filtered with a 100 Hz cutoff frequency using a 12 dB/octave Bessel filter (see insets) and the electromechanical shaker. (c),(d) Corresponding power spectra of the time domain traces shown in (a),(b), respectively. In panel (c) the mechanical resonance of the accelerometer can be seen at 10 kHz as well as mechanical resonances of the piezoelectric shaker at 11 kHz and 20 kHz. While a far slower mechanical resonance can be observed in panels (b),(d).

Previous experiments with the optomechanical accelerometer showed that locking a laser to the side of an optical resonance provided high precision when measuring the thermomechanical noise of the mechanical resonator, but that the dynamic range was very low due to the limited linear region of the cavity resonance [21]. This approach also required *a priori* knowledge of the local cavity resonance slope which can limit the resulting accuracy of the acceleration measurement. Additionally, the laser lock used in this previous work operated with a low-bandwidth feedback controller such that low-frequency motion is nulled out by the controller while motion outside of the controller bandwidth results in a change in the reflected light from the cavity. As a result, it was not possible to measure displacement of the mechanical resonator within these two frequency bands simultaneously. The electro-optic frequency comb readout overcomes these challenges as shown below.

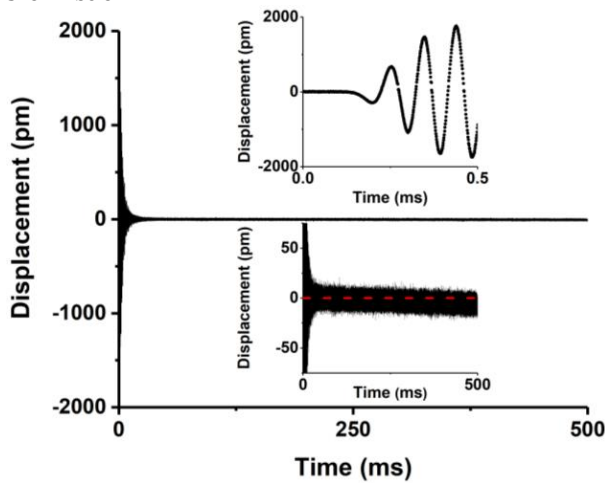

**Fig. 4.** Displacement of the mechanical resonator following excitation with a handheld force transducer. The accelerometer was mounted on a stainless-steel block with a low-friction bottom surface. The main panel shows the mechanical ring-down of the cavity due to the impulse. The upper inset shows the initial oscillations of the cavity following the impulse, which occurred at time zero. The measurement rate was 1 MHz, allowing for the rapid, large oscillations to be quantified with high fidelity. The lower inset reveals the slow near-DC cavity displacement resulting from the impact.

To further examine the dynamic measurement capabilities of the comb readout, two different types of shaker tables were used to excite the optomechanical accelerometer: a piezoelectric shaker table and an electromechanical shaker table with a voice coil actuator (subsequently referred to here as the mechanical shaker). First, we used the piezoelectric shaker to demonstrate the wide dynamic range of the electro-optic frequency comb readout method over four orders of magnitude of excitation, well beyond what is possible with a laser locking readout method.

The sensor was mounted on the shaker table and driven with a 4 kHz sinusoidal voltage at various amplitudes from 0.3 mV to 3 V. As can be seen in Fig. 2, the accelerometer is extremely linear over this very wide range with a standard deviation of the linear fit residuals of $9.6\times10^{-4}$ m/s$^2$ and a maximum deviation of only 0.4 % of the full measurement range. Though this data includes some contributions from nonlinearities in the piezoelectric actuator, we expect the shaker to be very linear over the small voltage range used here. We note the roll-off of this linearity plot at the lowest drive amplitudes is due to the presence of thermomechanical noise [21]. The maximum excitation voltage used here is 40 times greater than was possible with previous locking methods [21], demonstrating that electro-optic frequency comb readout can dramatically change what measurements are possible with cavity optomechanical systems. In addition, we note that the highest voltage utilized in this linearity measurement was limited by the available shaker drive source rather than the comb readout method which could record displacements a factor of five larger than those shown here.

In order to validate the measured accelerations we utilized a HeNe-based homodyne Michelson interferometer to measure the acceleration of a mirror mounted on the top of the accelerometer package. A further description of the homodyne interferometer can be found in Ref. [21] and its Supplemental Materials. As can be seen in Fig. 2 we observe excellent agreement between the accelerations measured with the comb-based-readout of the optomechanical sensor and the homodyne interferometer.

Next, we measured the accelerometer response to two different types of square-wave excitations. When mounted on the piezoelectric shaker table, we used a square-wave voltage with a period of 0.5 s to generate a fast step excitation and ring-down of the mechanical resonances, as shown in the time domain in Fig. 3a and frequency domain in Fig. 3c. The accelerometer resonance is visible near 10 kHz as well as the mechanical resonances of the shaker at 11 kHz and 20 kHz. Subsequently the accelerometer was placed on the mechanical shaker to examine its low frequency response. When the driving square-wave voltage was low-pass filtered at 100 Hz (i.e., well below the mechanical resonance of the accelerometer), a much slower time domain response was measured (see Fig. 3b and 3d) that clearly shows the change in direction of the step excitation within one period. See the Supplemental Materials for an additional example using a chirped sine-wave excitation.

As a demonstration of how comb readout can be used to measure large amplitude dynamic behavior at slow and fast time scales simultaneously, an impulse test was implemented. The accelerometer was mounted horizontally on a stainless-steel block that had a polytetrafluoroethylene coating on the bottom surface to provide low friction. The block was struck on the side opposing the accelerometer with a handheld force transducer which provided a convenient trigger signal.

Figure 4 shows the rapid, large cavity oscillations induced by the sudden impulse. During this initial impulse the cavity displacement is as large as 3.5 nm which corresponds to a motion of twenty-four cavity linewidths (1.8 GHz) with a period of only 100 µs (i.e., the period of the mechanical resonance). To the best of our knowledge, measurements with a laser locking system that can track this level of frequency change and slew rate have never been demonstrated. In addition to these rapid oscillations, the electro-optic comb readout also allows for the quantification of near-DC motion which is normally unobservable as it commonly lies within a cavity lock's servo bandwidth or requires a more difficult and harder-to-interpret in-loop measurement. In the lower inset we can observe a gradual displacement of the mechanical resonator over timescales out to 0.5 s.

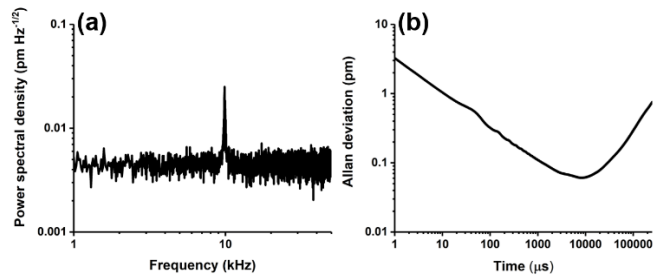

**Fig. 5.** (a) Noise power spectral density which shows the thermomechanical noise resonance near 10 kHz. (b) Corresponding overlapped Allan deviation [22, 23].

Finally, the limits of noise and stability were explored when the accelerometer was located on a vibration isolation platform within an acoustic enclosure. The power spectral density of the measured time-series data is shown in Fig. 5a, where the thermomechanical noise of the mechanical resonator is clearly observed and has an amplitude on resonance of 25 fm/√Hz which agrees with previous noise measurements of these devices [21]. The total optical power incident on the cavity was only 310 µW with each individual comb tooth having a power near 1 µW. Thus, the shown noise floor of 6 fm/√Hz was achieved with only 8 µW of incident optical power within the cavity's full-width at half-maximum which is forty-four times lower than the power previously utilized in the sidelock measurements [21]. The ability to operate at low intracavity power levels is advantageous to limit intracavity heating and thus bias instability in optical microcavities.

Using the same time series data, the overlapped Allan deviation [22, 23] was also calculated, as shown in Fig. 5b. This stability measurement is difficult to make with a traditional laser lock readout because the controller generally nulls out the quasi-static motion of the cavity or requires an in-loop measurement. Further, the controller can increase the instability due to added noise.

Here we have presented a new approach for the rapid and high-dynamic-range interrogation of optical cavities. Electro-optic frequency combs are particularly well suited to the readout of optomechanical devices and can simultaneously quantify a cavity's length, finesse, coupling efficiency and the presence of any interfering transverse modes—aspects of cavity optomechanical systems that are almost certainly varying dynamically but remain largely unexplored. Further, the use of a direct digital synthesis radiofrequency source leads to dramatically reduced cost and a small footprint. The described optics and electronics have been assembled in a portable rack-mounted system for mobile, robust operation. Importantly, the present approach does not suffer from many of the limitations present with laser locking approaches and is widely applicable for other physical and chemical measurements such as vibrometry and interferometry [24, 25].

**Funding and acknowledgements.** Y. B. was supported by the National Institute of Standards and Technology (70NANB17H247). In addition, the NIST on a Chip program provided partial support for this work. This research was performed in part at the NIST Center for Nanoscale Science and Technology NanoFab